\pdfoutput=1
\documentclass{article}
\usepackage{arxiv}
\usepackage[utf8]{inputenc} 
\usepackage[T1]{fontenc}    
\usepackage{hyperref}       
\usepackage{url}            
\usepackage{booktabs}       
\usepackage{amsfonts}       
\usepackage{nicefrac}       
\usepackage{microtype}      
\usepackage{lipsum}
\usepackage{natbib}
\usepackage{graphicx}
\usepackage{amsmath}
\usepackage{bm}
\usepackage{bbm}
\usepackage{mathrsfs}
\usepackage{subcaption}
\usepackage{epsfig}
\usepackage{amsfonts}
\usepackage{amssymb}
\usepackage{wrapfig}
\usepackage{psfrag}
\usepackage{hyperref}
\usepackage[ruled,commentsnumbered]{algorithm2e}
\usepackage{soul}
\usepackage{xcolor}
\newcommand{\gv}[1]{\ensuremath{\mbox{\boldmath$ #1 $}}} 
\newcommand{\grad}[1]{\gv{\nabla} #1} 
\newcommand{\red}[1]{\textcolor{black}{#1}}
\newcommand{\strike}[1]{}

\def\bI{\mbox{\boldmath$ I$}}

\def\bX{\mbox{\boldmath$ X$}}

\def\bc{\mbox{\boldmath$ c$}}

\def\bn{\mbox{\boldmath$ n$}}

\begin{document}

\title{A diffuse interface framework for modelling the evolution of multi-cell aggregates as a soft packing problem \red{driven by the} growth and division of cells}
\author{J. Jiang\thanks{Mechanical Engineering, University of Michigan}, K. Garikipati\thanks{Mechanical Engineering, and Mathematics, University of Michigan; corresponding author} \& S. Rudraraju\thanks{Mechanical Engineering, University of Wisconsin-Madison}}

\maketitle

\begin{abstract}
    We present a model for cell growth, division and packing under soft constraints that arise from the deformability of the cells as well as of a membrane that encloses them. Our treatment falls within the framework of diffuse interface methods, under which each cell is represented by a scalar phase field and the zero level set of the phase field represents the cell membrane. One crucial element in the treatment is the definition of a free energy density function that penalizes cell overlap, thus giving rise to a simple model of cell-cell contact. In order to properly represent cell packing and the associated free energy, we include a simplified representation of the anisotropic mechanical response of the underlying cytoskeleton and cell membrane through  penalization of the cell shape change. Numerical examples demonstrate the evolution of multi-cell clusters, and of the total free energy of the clusters as a consequence of growth, division and packing. 
\end{abstract}

\section{Introduction}
Formation of multi-cell aggregates is a foundational process in the evolution of multicellular organisms. Beginning with a single cell or a small cluster, the growth of aggregates is driven by cell division, differentiation, migration and cell-cell interactions. Understanding the processes underlying the formation of these aggregates is central to many phenomena in cellular biology and physiology, including embryogenesis, regeneration, wound healing, tissue engineering, and the growth and metastasis of cancerous tumors. As in most areas of cellular biology, a large body of work has focused on understanding the signalling pathways that control the evolution of multi-cell clusters, and it is widely accepted that these pathways are triggered by the chemical environment and mechanical interactions (intra-cell, cell-cell, aggregate-matrix and other external stimuli). However, the understanding of the spatial and temporal variations and effects of the chemo-mechanical environment in these cell aggregates remains at a nascent stage. Even in this early stage, because of its complexity, the field is increasingly leaning on computational models. 

Early work on modeling growth and interactions in cell clusters included lattice models \cite{GOEL1970423, GOEL1978103, Mochizuk1998}. These treat cells as sites on a square or hexagonal lattice and evolve multi-cell configurations through free energy-minimizing cell pair exchanges. These highly reduced order representations have provided significant insights into the effect of cell-cell interactions on the evolution of cell aggregates. However, the cell-cell exchange processes assumed in these models are not universally observed in real cell aggregates. Such treatments also limit the incorporation of sub-cellular growth dynamics. Improvement of the lattice models in the form of sub-cellular lattice models using high-Q Potts models have delivered better geometric representation of cell structure \cite{Glazier1992, Glazier1993}. In sub-lattice models, the cells are represented by a cluster of lattice sites rather than a single lattice site, and cell migration is achieved by switching parent cell identity of the lattice sites at the boundary. A single such switch allows the  
cell boundary of one cell to advance by one lattice length into the neighbouring cell. These methods have allowed for a finer representation of the cell geometry, but result in unrealistic jagged cell boundaries and complex cell shapes that are not simply connected (e.g. cells within cells). 

The drawbacks of the lattice models consisting of their non-representative cell geometries or a jagged representation of the cell boundaries was partially addressed with the development of cell-centric/center-dynamics models \cite{HONDA1978523, HONDA1983191, GRANER1993455, Mosaffa2015} and vertex dynamics models \cite{HONDA1983191, Honda1986, FLETCHER20142291, Silvanus2017, Munoz2017}. The center dynamics models approximate cell shapes as polygons generated though Voronoi tesselation of a collection of forming points, and the evolution of cell boundaries is achieved through a free energy minimizing movement of the forming points. A major drawback of this method is the restriction of the boundary to a polygonal shape dictated by the underlying tessellation. This restriction was addressed in vertex dynamics models that allowed for the cells to be represented as general polygons defined by the connectivity of the vertices. This connectivity evolves dynamically and is driven by the free energy minimizing pair wise movement of vertices that conserve the cell volume. 

As detailed in the review paper by Brodland \cite{Brodland2004}, the lattice models, cell-centric models and the vertex dynamics models have successfully modeled a wide range of cell-cell interactions and cell aggregates phenomena. However, these models have no representation or a very simplified representation of cell geometries, cell-cell and cell-substrate interactions, cytoskeletal remodeling, cytoplasmic viscosity, etc. This greatly limits the ability to model realistic, smooth and anisotropic cell shape evolution, the mechanics of cell surface evolution due to cell-cell contact, the ability to control cell volumes and to ensure proper dissipative dynamics. Modeling and understanding these processes is central to characterizing the process of growth and evolution of multi-cell aggregates that we refer to as a problem of soft packing.  

In this manuscript, we present a finite element method based phase field representation of cells and the resulting soft packing dynamics of cell aggregates. The phase field method is a popular numerical technique for simulating diffuse interface kinetics at the mesoscale and has been widely used to model evolving interface problems such as crystal growth, solidification and phase transformations in alloys. In this method, the evolution of a species concentration and/or phase is modelled using a set of conserved or non-conserved order parameters. The evolution of the order parameters and the corresponding interface kinetics are governed by a system of parabolic partial differential equations, which are referred to as the Cahn-Hilliard formulation (for conserved order parameters) \cite{Cahn1958} and Allen-Cahn formulation (for non-conserved order parameters)\cite{Cahn1979}. \red{The phase field representation of the cell geometries and the soft packing of multi-cell aggregates presented here allows for an improved representation of realistic cell shapes and the modeling of cell growth, division, and mechanical compaction processes intrinsic to the formation of multi-cell aggregates}. \red{\strike{Migration also can be treated in this setting, although it is beyond the scope of this communication.}} An earlier attempt at modeling multi-cell aggregates using a phase field representation was outlined by Nonomura \cite{Nonomura2012} using an Allen-Cahn representation of the cells; i.e. a non-conserved order parameter treatment. In contrast, the formulation presented in this paper treats cell mass as a conserved quantity and models the evolving cell clusters using a Cahn-Hilliard representation. Furthermore, our treatment considers the mechanics of soft packing and allows for the necessary anisotropic shape evolution of cells. 

In Section \ref{sec:nonmechformulation}, we present the phase field formulation, its numerical implementation and simulations of cell growth and division. This is followed by the modeling of mechanics of soft packing and simulations of soft packing in Section \ref{sec:mechformulation}. Closing remarks appear in Section \ref{sec:concl}.

\section{A phase field formulation for cell growth, division and contact}
\label{sec:nonmechformulation}
Our formulation of the problem rests on a phase field representation with as many scalar fields as cells. The treatment is centered on the definition of a free energy density, as a function of the scalar fields. In the non-mechanical version of the problem, terms are constructed to model cell membranes by phase segregation of cell interiors from the extra-cellular matrix, and contact inhibition, or intercellular repulsion, by penalizing overlapping scalar fields. We present the variational treatment, the mechanisms that model cell division, numerical aspects, and an illustrative numerical example.

\subsection{The diffuse interface model}
Let $\Omega\in \mathbb{R}^2$ with a smooth boundary $\partial \Omega$. Scalar fields $c_k, \;k = 1,\dots,N$ with $c_k \in [0,1]$ serve to delineate the interior and exterior of the cell numbered $k$. Here, the interior of cell $k$ is $\omega_k \subset \Omega$, where $\omega_k = \{ \bX \in \Omega\vert c_k(\bX) =1\}$. The exterior is $\Omega\backslash\omega_k$. The free energy density function is built up beginning with the following form: 
\begin{equation}
\psi_1(c_k) = \alpha c_k^2(c_k-1)^2 + \frac{\kappa}{2}\vert\grad c_k\vert^2
\label{equ:psi1}
\end{equation}
where the double-well term, $f(c_k) = \alpha c_k^2(c_k-1)^2$, enforces segregation into $\omega_k$ and $\Omega\backslash\omega_k$. In Equation \eqref{equ:psi1}, the second term enforces a diffuse cell-matrix interface (the cell membrane) of finite thickness, where $\kappa$ controls the interface thickness, and thereby the interfacial energy. For $N$ cells in $\Omega$, the above free energy density needs to be extended to model contact by adding a cell-cell repulsion term. The total free energy of the multi-cell aggregate is a functional $\Pi[\bc]$, defined as
\begin{align}
\Pi[\bc] &:= \int\limits_\Omega \psi(\bc,\grad c) ~\text{d}V\nonumber\\
&=\int\limits_\Omega \left(\sum_{k=1}^{N} f(c_k) + \sum_{k=1}^{N} \frac{\kappa}{2}\vert\grad c_k\vert^2 + \sum_{l\ne k}\sum_{k=1}^{N}\lambda c_k^2 c_l^2 \right) ~\text{d}V.
\label{equ:energy}
\end{align}
Here, $\bc = \{c_1,\dots,c_N\}$, and $\lambda$ is a penalty coefficient that enforces repulsion between any two cells $k,l$ thus modelling cell contact.

Taking the variational derivative with respect to $c_k$ in Equation \eqref{equ:energy} yields
\begin{align}
\delta \Pi_k[\bc;w] =  & \left.\frac{\text{d}}{\text{d}\epsilon} \int\limits_{\Omega} \sum_{k=1}^{N} \left(f(c_k+\epsilon w) + \frac{\kappa}{2}\vert\grad (c_k+\epsilon w)\vert^2+ \sum_{l\ne k}\lambda (c_k+\epsilon w)^2 c_l^2\right)  ~\text{d}V \right|_{\epsilon=0} \nonumber\\
=&\int\limits_{\Omega} w \left( f^\prime(c_k) -  \kappa \Delta  c_k  + \sum_{l\ne k}2\lambda  c_k c_l^2 \right) ~dV + \int\limits_{\partial \Omega}   w \kappa \grad c_k \cdot \bn   ~dS
\end{align} 
where $\bn$ is the unit outward normal vector to $\partial \Omega$. The chemical potential of the $k^\text{th}$ cell is identified as,
\begin{equation}
\mu_k  = f^\prime(c_k) -  \kappa \Delta c_k + \sum_{l\ne k}2\lambda  c_k c_l^2
\label{equ:chemo_potential}
\end{equation}
At equilibrium, $\delta_k \Pi[\bc;w] =0$ for the $k^\text{th}$ cell, yielding $\mu_k = 0$ in $\Omega$, and $\kappa \grad c_k \cdot \bn = 0$ on $\partial \Omega$. \footnote{In some simulations, a buffer zone, $\Omega^{'}$, is needed around the simulation domain to inhibit unrealistic cell shapes resulting from the enforcement of $\kappa \grad c_k \cdot \bn = 0$ on $\partial \Omega$. In addition, this buffer zone also acts as a membrane around the cell cluster. In the buffer zone, an additional term of the form $\sum_{k=1}^{N}\lambda c_k^2$ is added to the free energy density to penalize the movement of any cells from the active simulation domain ($\Omega$) to the buffer zone ($\Omega^{'}$).} 

The following parabolic partial differential equation, popularly known as the Cahn-Hilliard equation \cite{Cahn1958}, imposes the conserved dynamics that governs the delineation and growth of the $N-$cell agglomerate, and of repulsion between cell pairs:
\begin{equation}
\frac{\partial c_k}{\partial t} = -~\grad \cdot (-M\grad \mu_k) + s_k
\label{equ:dynamic}
\end{equation}
where the source term $s_k$ has been introduced, and $M$ is the mobility, assumed to be constant. The dynamics of the multi-cell soft packing problem is governed by Equation \eqref{equ:dynamic} with the thermodynamics prescribed by Equation \eqref{equ:chemo_potential} and boundary condition $\kappa \grad c_k \cdot \bn = 0$ on $\partial\Omega$ for $k = 1,\dots N$.

\subsection{Numerical implementation}
\label{sec:chemo_numerical}
Time discretization is carried out by the implicit, backward Euler method. Time instants are indexed by superscripts in the following development, and the time step is denoted by $\Delta t = t^{n+1}-t^n$. Starting with the initial conditions $\{c_k^0,\mu_k^0\}$, and given the solution $\{c_k^n,\mu_k^n\}$ at time $t^n$, the time-discrete versions of Equations \eqref{equ:dynamic} and \eqref{equ:chemo_potential} are,
\begin{align}
c^{n+1}_{k} &= c^{n}_{k} + \Delta t (M~\grad \cdot (\grad \mu^{n+1}_{k})+s_{k})\nonumber\\
\text{where}\quad\mu^{n+1}_{k} &= {f^\prime}^ {n+1}(c_k) -  \kappa \Delta c^{n+1}_{k} + \sum_{l\ne k}2\lambda  c_k^{n+1} {c_l^{n+1}}^2
\label{equ:time_discretization}
\end{align}

\red{The weak form, posed as a classical two-field mixed finite element formulation \cite{Brezzi1991}, is stated as follows: Find $c^{n+1}_k \in \mathscr{S} = \{c \vert c \in \mathscr{H}^1(\Omega),  \grad c \cdot\bn = 0 \;\text{on}\; \partial\Omega \}$ \footnote{\red{In this initial boundary value problem, $\partial\Omega$ is composed of only a Neumann (flux) boundary ($\partial\Omega^{h} = \partial\Omega$) and hence the Dirichlet boundary is a null set ($\partial\Omega^{g} = \varnothing$).}} and $\mu^{n+1}_k \in \mathscr{S} = \{\mu \vert \mu\in \mathscr{H}^1(\Omega)\}$ such that for all variations $w \in \mathscr{V} = \{w \vert w\in \mathscr{H}^1(\Omega)\}$ on $c_k$ and $v \in \mathscr{V} = \{v \vert v\in \mathscr{H}^1(\Omega)\}$ on $\mu_k$, respectively, the following residual equations  are satisfied:
\begin{align}
\int\limits_{\Omega}   w c_k^{n+1} ~\text{d}V
&= \int\limits_{\Omega}   (w  c_k^{n} - \grad w \cdot \Delta t M \grad \mu_k^{n+1} + w\Delta t s_k )~\text{d}V\nonumber\\
\int\limits_{\Omega}   v  \mu_{k}^{n+1}  ~\text{d}V 
&=\int\limits_{\Omega}  (v  {f^\prime}^{n+1}(c_k) + \grad v \cdot \kappa \grad c_k^{n+1} )~dV + \int\limits_\Omega v \sum_{l\ne k}2\lambda  c_k^{n+1} {c_l^{n+1}}^2  ~\text{d}V
\label{equ:weak}
\end{align}
}
Spatial discretization is implemented in a standard finite element framework and uses bilinear quadrilateral elements leading to standard matrix-vector forms of the equations in \eqref{equ:weak}.

\subsection{The model for cell division}
\label{subsec:cellDivision}
\red{Two criteria of cell division are considered:  Age-based and mass-based. In the age-based criterion, the cells divide after reaching a specified physical age irrespective of their current mass. In the mass-based criterion, the cells divide after reaching twice their mass at birth irrespective of the time it takes to reach that mass. If the cell division process is spatially symmetric and leads to bisection of the parent cell, and if the daughter cells have the same growth rate, then both these criteria can lead to a sequence of divisions at the same time instants. However, asymmetrical cell division is common due to small perturbations, which, in our model, are introduced by small numerical differences. As a result, different sequences of evolution of the cell clusters can result with the two criteria. Furthermore, the distinction introduced by absence/presence of the source term, $s_k$, with the time-/mass-based criteria, respectively, in our implementation (see Section \ref{sec:const-varyingvolume}), contributes more significant differences. The implementation of these criteria is outlined in Algorithm \ref{Algo:division}, and the process of cell division proceeds as follows:} For an elliptical cell dividing along its minor principal axis, $\omega_{N+1}^{n+1}$ is defined such that $\omega_k^{n}$ is bisected into $\omega_k^{n+1}$ and $\omega_{N+1}^{n+1}$, ensuring $\text{meas}(\omega_{k}^{n+1}) = 0.5\text{meas}(\omega_k^{n})$ and $\text{meas}(\omega_{N+1}^{n+1}) = 0.5\text{meas}(\omega_k^{n})$, where $\text{meas}(\omega_k)$ is the volume (area in two dimensions) of $\omega_k$.
The regions $\omega_k^{n+1}$ and $\omega_{N+1}^{n+1}$ are thus determined by the divisions of elliptical cells along their minor principal axes. This strategy of bisecting an elliptical cell along its minor axes is motivated by observations of cell division in biological cells. However, recognizing that the shape of $\omega_k^n$ will, in general deviate from an ellipse, we define the division axis to lie along the axis of the major principal moment of inertia through the center of mass. Note that, for an ellipse, the minor principal axis, and the major principal moment of inertia axis through the center of mass coincide. At time $t^{n+1}$, a new interface forms between $\omega_k^{n+1}$ and $\omega_{N+1}^{n+1}$ following a division of the $k^\text{th}$ cell at time $t^n$ that incremented the number of cells $N\mapsto N+1$. This new interface is along the major principal axis of the moment of inertia tensor through the center of mass of $\omega^n_k$; that is, of the $k^\text{th}$ cell's interior at time $t^n$. \\
\red{Numerical implementation of the division process involves introducing two new fields, $c_{N+1}$ and $\mu_{N+1}$, over the problem domain to represent the new cell $\omega_{N+1}^{n+1}$. This is followed by initializing $c_{N+1}$ and resetting $c_k$ as follows: 
\begin{flalign*}
c^{n+1}_{N+1}(\bX) &=c^{n}_{k}(\bX)  ~\forall ~\bX \in \omega^{n+1}_{N+1} &&\\
c^{n+1}_{N+1}(\bX) &=0.0  \hskip1.5em\relax ~\forall ~\bX \in \omega^{n+1}_{k} &&\\
c^{n+1}_{k}(\bX) &=0.0  \hskip1.5em\relax ~\forall ~\bX \in \omega^{n+1}_{N+1} &&
\end{flalign*}
So computationally, during every cell division, two new fields are introduced, and the total number of degrees of freedom in the problem is increased by $2N_{nodes}$, where $N_{nodes}$ is the number of nodes/mesh-points in the discretization of the problem domain. This would cause the number of degrees of freedom to linearly increase with the number of cells, becoming computationally expensive for simulations with large clusters of cells. However, this limitation can be eliminated by  active parameter tracking methods developed in the phase field community \cite{vedantam2006}. These methods leverage graph theory to introduce ``coloring schemes'' that can represent arbitrary numbers of cells with a finite number of fields (typically less than 10). This is accomplished by restricting the support of each field to a neighbourhood that is only slightly larger than the corresponding cell, instead of the entire problem domain.}

\begin{algorithm}[h]
\vspace{0.1in}
\red{
\underline{Age based division criterion:}\\ \vspace{0.025in}
$T^{\text{age}}_k$ is the current age of the cell $\omega_k$,
and $T^{\text{division}}_k$ is the age at division.\\
\If{$\varepsilon_1 < T^{\text{division}}_k/T^{\text{age}}_k -1 < \varepsilon_2$}{
    $N\mapsto N+1$\;
    $\text{meas}(\omega_{k}^{n+1}) = 0.5\text{meas}(\omega_k^{n})$\;  $\text{meas}(\omega_{N+1}^{n+1}) = 0.5\text{meas}(\omega_k^{n})$\;
}}
\vspace{0.1in}
\underline{Mass based division criterion:}\\ \vspace{0.025in}
$m_k^{n}$ is the current mass of the cell $\omega_k$, given by: $m_k^{n} = \int_\Omega c_k^{n} ~dV$\;
\If{$\varepsilon_1 < m_k^n/2m_k^0 -1 < \varepsilon_2$}{
    $N\mapsto N+1$\;
    $\text{meas}(\omega_{k}^{n+1}) = 0.5\text{meas}(\omega_k^{n})$\;  $\text{meas}(\omega_{N+1}^{n+1}) = 0.5\text{meas}(\omega_k^{n})$\;
}
\caption{Cell division mechanisms, given $0< \varepsilon_1 < \varepsilon_2 \ll 1$}
\label{Algo:division}
\end{algorithm}

\subsection{Adaptive time stepping}
A uniform time step, $\Delta t = \overline{\Delta t}$, is chosen such that it ensures the stability and convergence of the Cahn-Hilliard dynamics. However, when cell divisions occur, adaptive time step control is necessary to equilibrate the large penalty forces arising from sharply varying composition fields over the new daughter cells' common boundary. To address the associated, transient numerical stiffness, $\Delta t$ is decreased by a factor of $1.0\times 10^{-m}$ for a few time steps ($n_\text{div}$) in order to ensure convergence of the nonlinear system of equations. The time step is then sharply increased back to $\Delta t = \overline{\Delta t}$ until the next cell division process. Algorithm \ref{Algo:Timestep} details the implementation of this adaptive time step control in our code. Typical values used for the simulations presented in this paper are $n_{div}<5$ and $m<7$.

\begin{algorithm}[h]
$\Delta t=\overline{\Delta t}$\;
\For{$t\leftarrow 0$ \KwTo T ; $t^{n+1}=t^n+\Delta t$}{
    \tcc{T is the total time of computation; $\Delta t$ is the time step}
    \If{cell is about to divide}{
        $\Delta t=\overline{\Delta t}\times 1.0\times 10^{-m}$\;
        $a=0$\;
        \If{$\Delta t<\overline{\Delta t}$}{
            $a=a+1$\;
            \If{$a>n_{div}$}{
                $\Delta t=\overline{\Delta t}\times 1.0\times 10^{-m}\times(a-n_{div})^m$\;
            }
        }
    }
}
\caption{Adaptive time step control\label{Algo:Timestep}}
\end{algorithm}

\subsection{Adaptive mass source}
\label{sec:source}
The growth of cells is controlled by the source term, $s_k$, which can be determined from experimentally observed cell doubling time estimates. In this work, $s_k$ is a function of the mass ratio $\nu_k^n = m_k^n/m_k^0$: The default value of $s_k=\overline{s}$, the average growth rate. Growth continues at this rate until the cell has doubled in mass and then the division process separates the cell into two daughter cells. The two daughter cells then continue to grow at the rate $\overline{s}$. However, the total number of cells in the limit of optimal soft packing is given by $N_\text{max}=\text{meas}(\Omega)/m_k^0$. If either the number of cells approaches $N_\text{max}$ or if no new scalar fields are available to initialize new daughter cells leading to $\nu_k^n > 1.1$, we turn off the source, $s_k^{n+1} = 0$, so that cell $k$ no longer grows. This adaptive mass source control appears in Algorithm \ref{Algo:Source}.

\begin{algorithm}[h]
\If{$\bX \in \omega^n_k$; in cell $k$}{
    $s^n_k = \overline{s}$\;
    $\nu^n_k = m^n_k/m^0_k$\;
    \tcc{ratio of current cell mass to its initial mass}
    \If{$N \ge N_\text{max}$; new phase field is not available}{
        \If{$\nu^n_k > 1.1$}{
            $s^n_k = 0.0$\;
        }
    }
    
}
\caption{Adaptive mass source\label{Algo:Source}}
\end{algorithm}

\subsection{Code framework}
The two-dimensional, cell growth and soft packing formulation presented here has been implemented in the \texttt{C++} based \texttt{deal.II} open source parallel finite element library \cite{BangerthHartmannKanschat2007}. We use the \texttt{SuperLU} direct solver \cite{li05} to solve the system of linear equations obtained from the linearization of Equation \eqref{equ:weak}. The linearization itself is obtained using the \texttt{Sacado} algorithmic differentiation library of the open source \texttt{Trilinos} project \cite{Trilinos2003}.  


\subsection{Illustrative examples for the progression of cell growth, division and compaction.}
\label{sec:const-varyingvolume}
\begin{figure}[ht]
	\centering
	\begin{subfigure}[h!]{0.27\textwidth}
		\includegraphics[width=\textwidth]{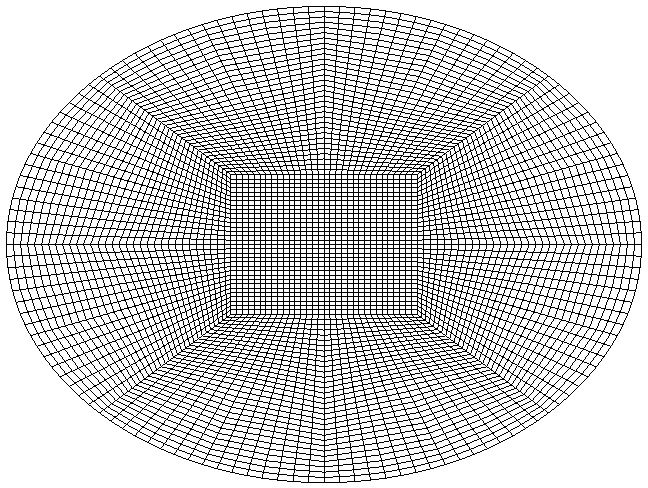}
		\caption{Spatial discretization.}
		\label{fig:mesh}
	\end{subfigure}
	\begin{subfigure}[h!]{0.27\textwidth}
		\includegraphics[width=\textwidth]{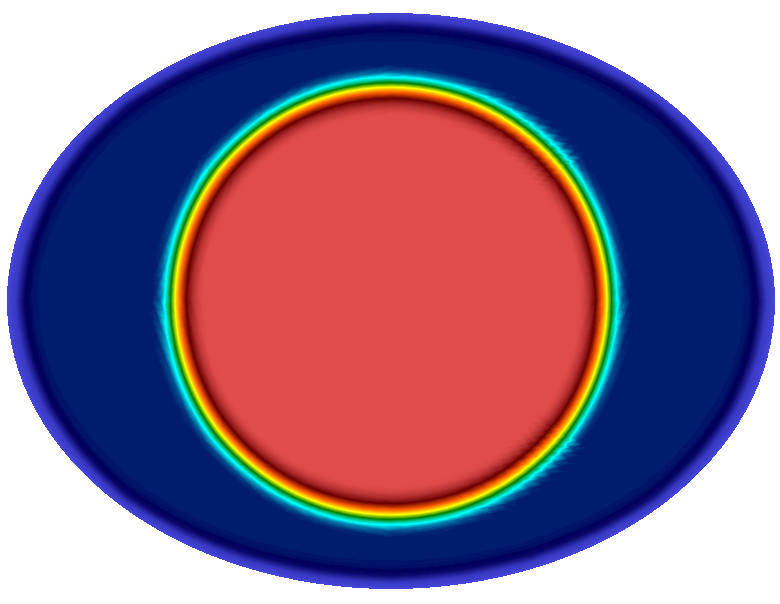}
		\caption{Initial circular cell.}
		\label{fig:16cells0}
	\end{subfigure}
	\begin{subfigure}[h!]{0.27\textwidth}
		\includegraphics[width=\textwidth]{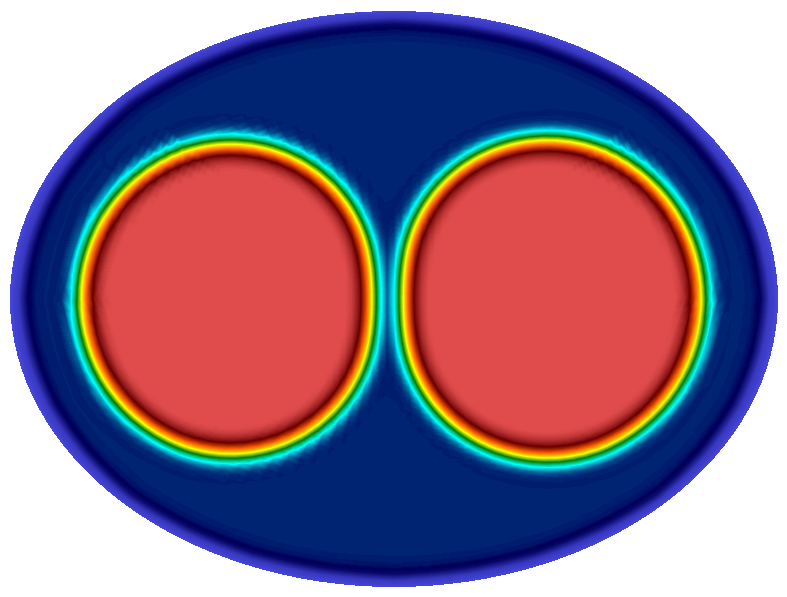}
		\caption{Division to two cells.}
		\label{fig:16cells1}
	\end{subfigure}
	\begin{subfigure}[h!]{0.27\textwidth}
		\includegraphics[width=\textwidth]{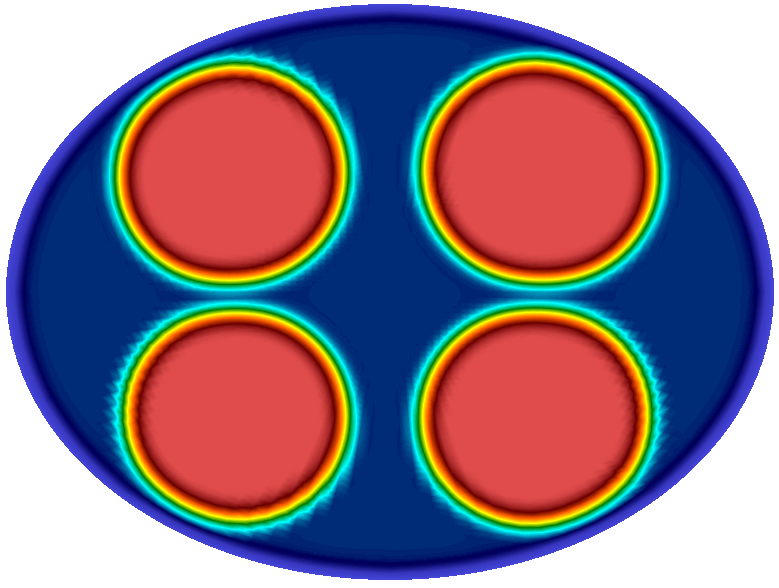}
		\caption{Division to four cells.}
		\label{fig:16cells2}
	\end{subfigure}
	\begin{subfigure}[h!]{0.27\textwidth}
		\includegraphics[width=\textwidth]{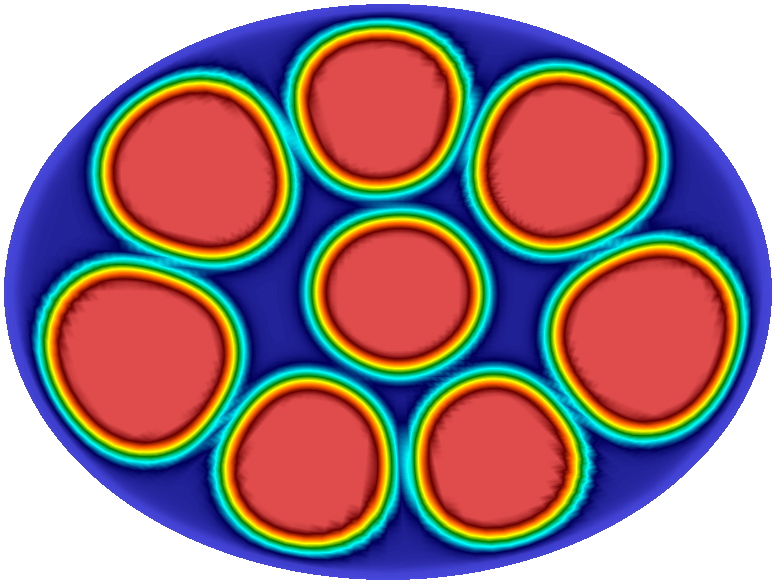}
		\caption{Division to eight cells.}
		\label{fig:16cells2_2}
	\end{subfigure}
		\begin{subfigure}[h!]{0.27\textwidth}
		\includegraphics[width=\textwidth]{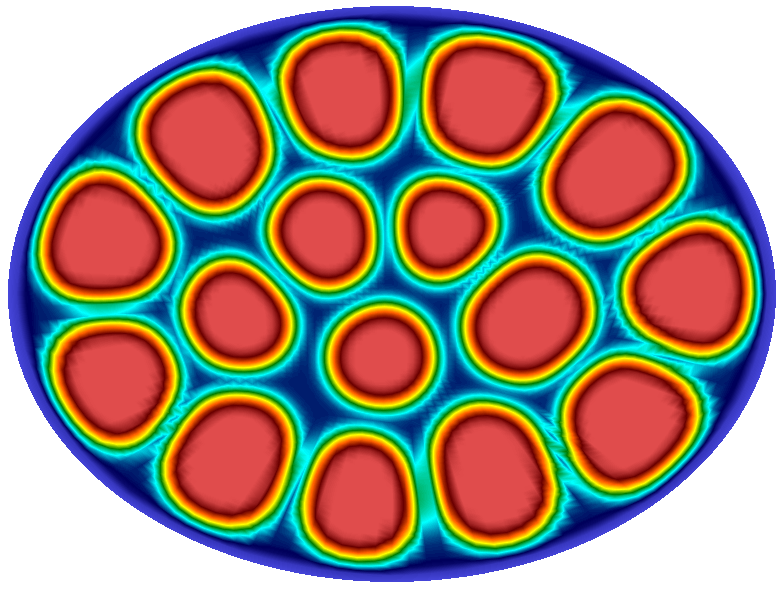}
		\caption{Division to sixteen cells and compaction.}
		\label{fig:16cells4}
	\end{subfigure}
	\caption{\red{A demonstration of the progression of cell division from one cell into sixteen cells at a constant total cell volume. The underlying spatial mesh/discretization over the simulation domain is shown in (a). Cell interiors are shown in red and the cell membrane in cyan-yellow.}}
	\label{fig:celldivisionConstVolume}
\end{figure}

\begin{figure}[ht]
	\centering
	\begin{subfigure}[h!]{0.35\textwidth}
		\includegraphics[width=\textwidth]{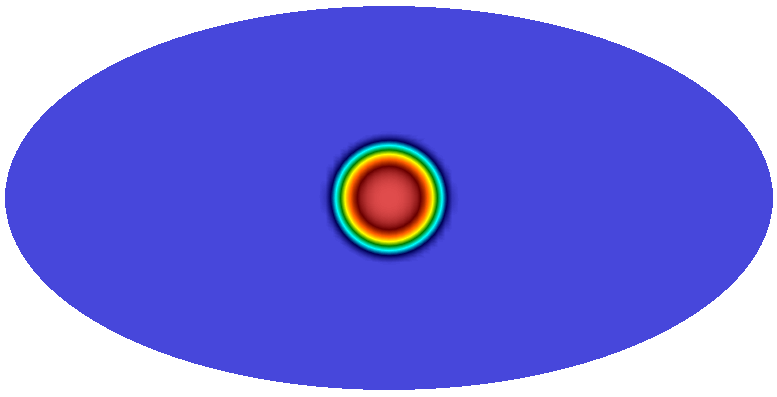}
		\caption{Initial circular cell.}
		\label{fig:onecell}
	\end{subfigure}
	\begin{subfigure}[h!]{0.35\textwidth}
		\includegraphics[width=\textwidth]{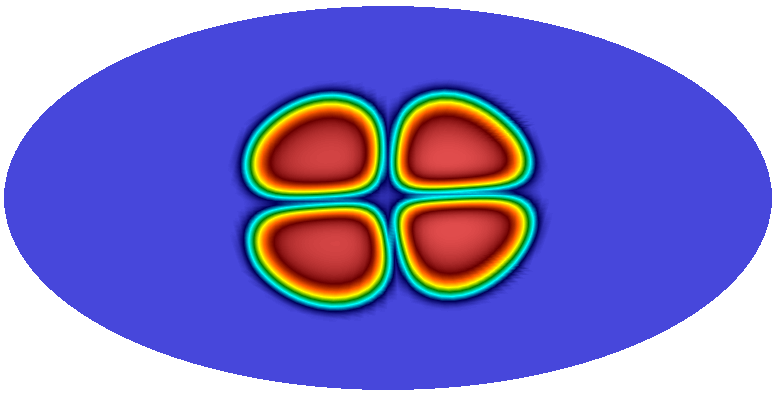}
		\caption{Progression to four cells.}
		\label{fig:fourcell}
	\end{subfigure}
	\begin{subfigure}[h!]{0.35\textwidth}
		\includegraphics[width=\textwidth]{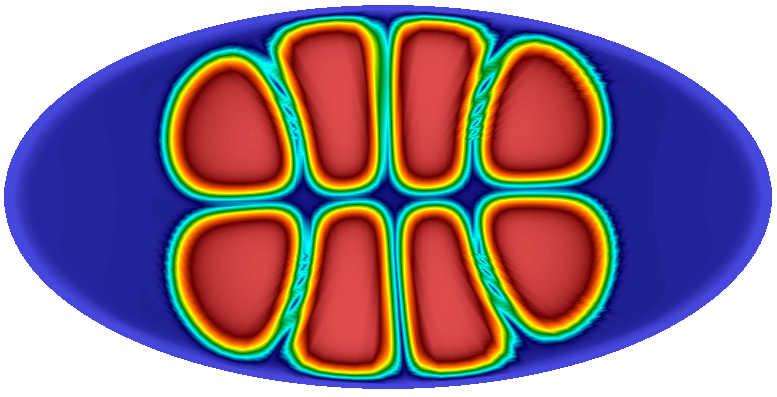}
		\caption{Progression to eight cells.}
		\label{fig:eightcell}
	\end{subfigure}
	\begin{subfigure}[h!]{0.35\textwidth}
		\includegraphics[width=\textwidth]{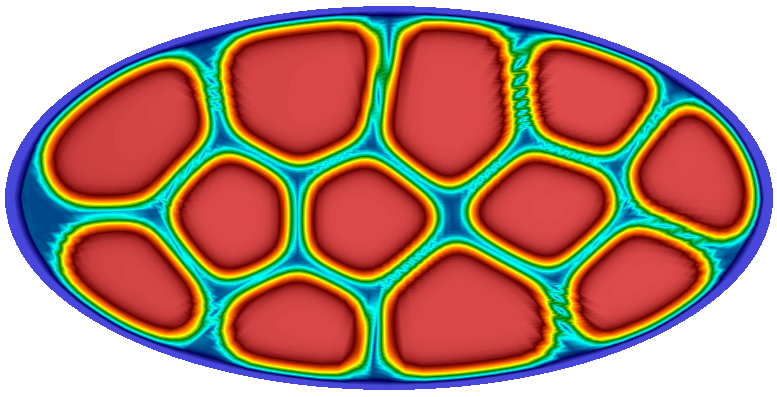}
		\caption{Progression to twelve cells.}
		\label{fig:twelvecell}
	\end{subfigure}
	\caption{A demonstration of the progression of cell division and growth from one cell into twelve cells driven by a source term. Cell interiors are shown in red and the cell membrane in cyan-yellow.}
	\label{fig:celldivision}
\end{figure}

\begin{figure}[ht]
\centering
\includegraphics[scale=0.5]{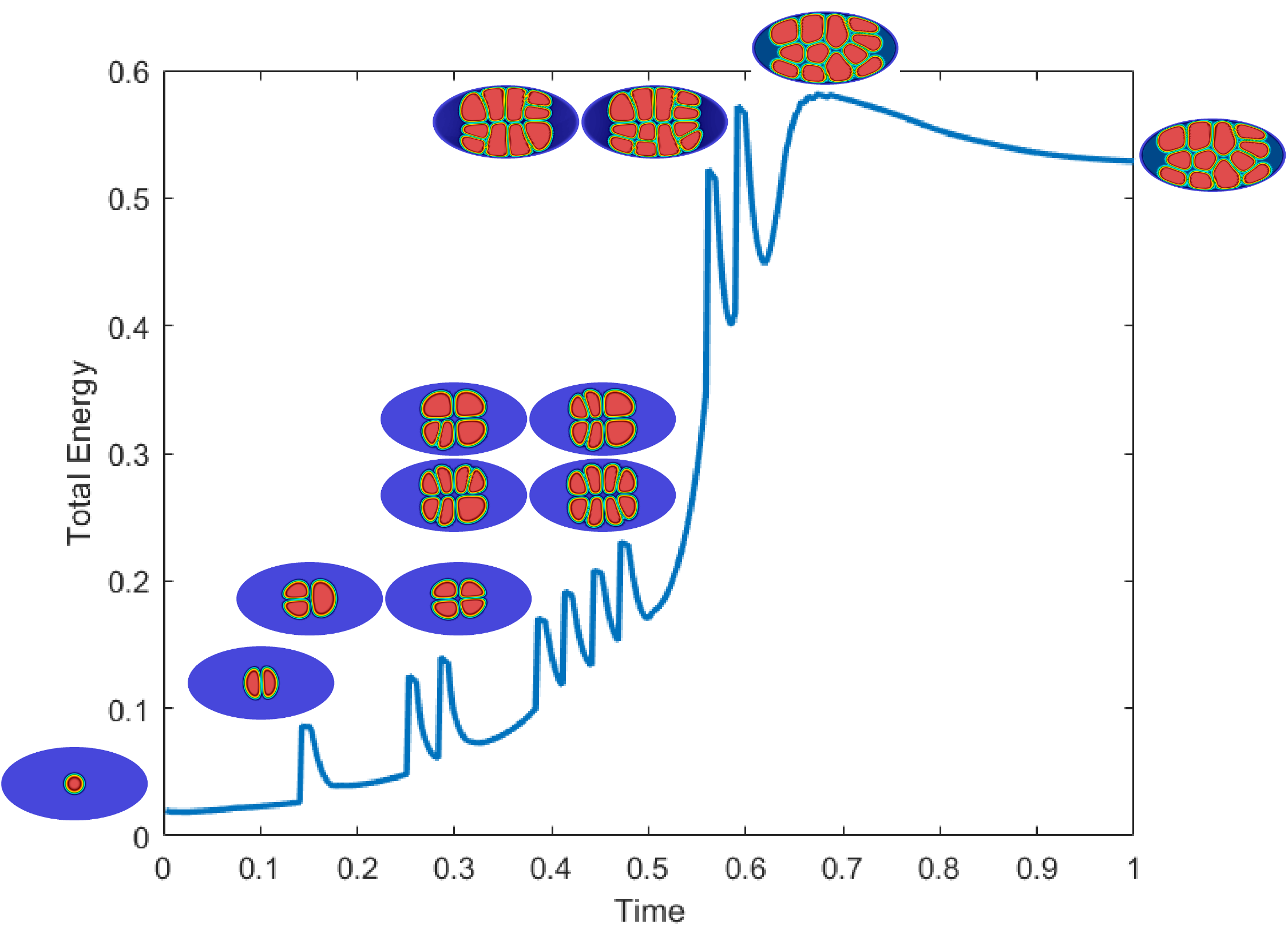}
\caption{Evolution of total free energy with time (normalized). \red{Each spike in the energy curve corresponds to transient repulsion between newly formed daughter cells following a cell division. The corresponding cell division events are shown in the inset sub-figures.}}
\label{fig:energy}
\end{figure}

We consider two cases for demonstrating the evolution of cell growth, division and contact leading to compaction:
\red{
\begin{itemize}
\item \emph{Cell division at a constant volume}, $s_k = 0$:  Early stages of embryonic cell division in many organisms (e.g. Caenorhabditis elegans \cite{Gilbert2000}) is known to occur at a constant embryonic volume. During these early stages, there is an increase in cell numbers, but the overall embryo volume stays fixed. This process is modeled in Figure \ref{fig:celldivisionConstVolume}, where a single circular cell divides into $N_\text{max} = 16$ cells while maintaining the total cell volume fixed. Here we use the age based criterion for cell division as the mass of each daughter cell does not change over time.
\item \emph{Cell division with a source}, $s_k \neq 0$: In species where the early embryo grows in size, the total cytoplasmic volume of all the cells increases with time. This process of growth of the embryonic cells can be modeled via a positive source term in Equation \eqref{equ:dynamic}. The corresponding simulation for $N_\text{max} = 12$ is shown in Figure \ref{fig:celldivision}. Here we use the mass based criterion for cell division as the positive source term leads to growth of the daughter cells.
\end{itemize}
The parameters used in these numerical examples appear in Table \ref{tab:parameter}. Our computations begin with a single cell, $\omega_1^0$, of circular shape at the center of an elliptical ``embryo'' $\Omega$.  Each division axis is determined by the major principal direction of the moment of inertia tensor through the center of mass, which distributes mass evenly in each daughter cell. This can be observed from the progression in Figures \ref{fig:16cells0}--\ref{fig:16cells4} for division at constant total cell volume, and in Figures \ref{fig:onecell}--\ref{fig:twelvecell} for division and growth. Attention is also drawn to the delineation of daughter cells following each division, and of non-sibling cells from each other due to the repulsion terms in Equation \eqref{equ:energy}.}  \\
Figure \ref{fig:energy} tracks the total free energy of the system for the case of cell division and growth with a source shown in Figure \ref{fig:celldivision}. Note that the total free energy increases with time due to the mass supply, and that transient fluctuations occur at each cell division due to the formation of a sharp boundary between the daughter cells and the transiently stronger repulsion between them. \red{When the source term vanishes (after the last division and corresponding energy spike), the free energy monotonically decreases, as dictated by the dissipative nature of phase field dynamics.} The initial conditions for the phase field for the computations in Figures \ref{fig:celldivisionConstVolume} and \ref{fig:celldivision} include a small random perturbation from a mean value to help drive the Cahn-Hilliard dynamics. This perturbation results in a tilt in the major principal direction of the moment of inertia tensor, and asymmetric division of the cells ($\sim 0.05$ radians off the axes of symmetry).\\
\red{The computations in Figures \ref{fig:celldivisionConstVolume} and \ref{fig:celldivision} were performed on an elliptical domain (mesh shown in Figure \ref{fig:mesh} for the constant volume case) with 5120 elements and 5185 nodes. Since each cell is represented by two scalar fields, $c$ and $\mu$, the degrees of freedom associated with each cell are 10370. For the simulation shown in Figure \ref{fig:celldivisionConstVolume} involving sixteen cells, this resulted in 165,920 total degrees of freedom.} 

\begin{table}[ht]
	\caption{Numerical values of parameters}
	\begin{center}
		\begin{tabular}{|c|c|c|c|c|c|c|}
			\hline
			Parameters & $\alpha$ & $\kappa$ & $\lambda$ & $M$ & $s$ & $\overline{\Delta t}$\\
			\hline
			Constant volume & $4$ & $1.0\mathrm{e}{-3}$ & $100$ & $1$ & 0.0 & $2.0\mathrm{e}{-4}$\\
			\hline
			With source & $4$ & $1.0\mathrm{e}{-3}$ & $100$ & $1$ & $5.0\mathrm{e}2$ & $2.0\mathrm{e}{-4}$\\
			\hline
		\end{tabular}
	\end{center}
	\label{tab:parameter}
\end{table}

\section{Mechanics of soft packing, driven by cell shape changes}
\label{sec:mechformulation}
We now present an extension of the formulation to an elementary mechanical model that associates energy to global shape changes of a cell. The model works by penalizing departures of the principal moments of inertia from their initial values, which are established when a specific cell is created. A more complete treatment of mechanics in the cell growth, division and soft packing problem would consider nonlinearly elastic interactions between cells. However, as we demonstrate, the simple shape change-based model does capture the essential mechanics of soft packing. We include a study of the reduced moduli that govern shape change via a parametric study, before studying a sequence of cell growth, divisions and soft packing with the added mechanical contributions.

\subsection{An extension of the free energy density function to incorporate the mechanics of cell shape change}
Consider the addition of the following term to the free energy density \eqref{equ:energy} in order to account for the principal moments of inertia of a single cell: 
\begin{equation}
\Pi_{\text{MI}} = \sum_{i=1}^{\text{dim}}\delta^i ({I^i} ^\text{ref}-I^i)^2
\label{equ:MI}
\end{equation}
Here,  $\delta^i$ is a mechanical modulus penalizing variations in the $i^\text{th}$ principal moment of inertia, $I^i$, from its  reference state, ${I^i}^\text{ref}$, established at the birth of this cell. With this addition that accounts for changes in shape, the total free energy function takes on the form:
\begin{align}
\Pi[\bc,\grad \bc] &:= \int\limits_{\Omega} \left(\sum_{k=1}^{N} f(c_k) + \sum_{k=1}^{N} \frac{\kappa}{2}\vert\grad c_k\vert^2 + \sum_{l\ne k}\sum_{k=1}^{N}\lambda c_k^2 c_l^2 \right) ~\text{d}V\nonumber\\
&\phantom{+}+ \sum_{k=1}^{N}\sum_{i=1}^{\text{dim}}\delta_k^i ({I_k^i} ^\text{ref}-I_k^i)^2
\end{align}
The moment of inertia tensor through the center of mass, $\bI$, is expressed in terms of its Cartesian components $I_{ij}$, with the phase field playing the role of the mass density in the traditional definition of this quantity:
\begin{equation}
\bI[c] = 
\begin{bmatrix}
I_{11}[c] & I_{12}[c] \\
I_{21}[c] & I_{22}[c]
\end{bmatrix}
=
\begin{bmatrix}
\int c \bar{X}_2^2 ~\text{d}V   & -\int c\bar{X}_1 \bar{X}_2 ~\text{d}V\\
-\int c\bar{X}_1 \bar{X}_2 ~\text{d}V & \int c\bar{X}_1^2 ~\text{d}V
\end{bmatrix}
\label{equ:MomentOfInertia}
\end{equation}
where, with the center of mass
\begin{equation}
    X^c_i = \frac{\int_\Omega c X_i ~\text{d}V}{\int_\Omega c ~\text{d}V},
    \label{equ:masscenter}
\end{equation} the Cartesian coordinates relative to the center of mass are $\bar{X}_{i} = X_{i}-X^c_{i}$. Then, the principal moments of inertia through the center of mass, $I^i$, can be determined from the moment of inertia tensor in Equation \eqref{equ:MomentOfInertia} by an eigen decomposition governed by the Cayley-Hamilton Theorem:
\begin{equation}
    I_k^2 - I_k\text{tr}\bI_k + \text{det}\bI_k = 0
    \label{equ:CHThm}
\end{equation}
where $I_k$ denotes a principal value of the moment of inertia tensor, relative to the center of mass, of the $k^\text{th}$ cell, $\bI_k$.

The variational machinery applied to computing the chemical potential now must account for the added, shape-dependent, moment of inertia terms:
\begin{align}
\delta \Pi_k[\bc;w]=  &\frac{\text{d}}{\text{d}\epsilon} \int\limits_{\Omega} \sum_{k=1}^{N}\left( f(c_k+\epsilon w) + \frac{\kappa}{2}\vert\grad (c_k+\epsilon w)\vert^2 + \sum_{l\ne k}\lambda (c_k+\epsilon w)^2 c_l^2 \right) ~\text{d}V \nonumber\\
&+ \frac{\text{d}}{\text{d}\epsilon}\left. \sum_{k=1}^{N}\sum_{i=1}^{\text{dim}}\delta_k^i \left({I_k^i} ^\text{ref}-I^i_k[c_k+\epsilon w]\right)^2 \right\vert_{\epsilon=0} \nonumber\\
= &\sum_{k=1}^{N} \int\limits_{\Omega} w \left( {f^\prime}(c_k) -  \kappa \Delta  c_k  + \sum_{l\ne k}2\lambda  c_k c_l^2 \right) ~\text{d}V \nonumber \\
&- \sum_{k=1}^N\sum_{i=1}^{\text{dim}} 2\delta_k^i \left({I_k^i} ^\text{ref} - I_k^i[c_k]\right)\tilde{I}^i_k[c_k] \nonumber	\\
&+ \int\limits_{\partial \Omega}   w \kappa \grad c_k \cdot \bn   ~dS
\end{align}
where, the term $\tilde{I}^i_k$ arises from the variation of $I^i_k$, and is obtained from the Cayley-Hamilton Theorem to be:

\begin{equation}
    \tilde{I}^i_k = \frac{\int\limits_\Omega w\left(I^i_k\text{tr}\bar{\bI}-\bar{X}_1^2\int\limits_\Omega c\,\bar{X}_2^2\,\text{d}V -\bar{X}_2^2\int\limits_\Omega c\,\bar{X}_1^2\,\text{d}V+2\bar{X}_1\bar{X}_2\int\limits_\Omega c\,\bar{X}_1\bar{X}_2\text{d}V\right)\text{d}V}{2 I^i_k - \text{tr}\bI},
    \label{equ:CayleyHamilton}
\end{equation}
with the tensor 
\begin{equation}
    \bar{\bI} = 
    \begin{bmatrix}
\bar{X}_2^2   & -\bar{X}_1 \bar{X}_2\\
-\bar{X}_1 \bar{X}_2 & \bar{X}_1^2
\end{bmatrix}.
\label{equ:barI}
\end{equation}

Extending Equation \eqref{equ:chemo_potential}, the chemical potential is now defined as
\begin{equation}
\mu_k  = f^\prime(c_k) -  \kappa \Delta c_k + \sum_{l\ne k}2\lambda  c_k c_l^2 - \sum_{i=1}^{\text{dim}} 2\delta_k^i \left({I_k^i} ^\text{ref} - I_k^i\right)\hat{I}^i_k,
\label{equ:chempotmech}
\end{equation}
where $\hat{I}^i_k$ is
\begin{equation}
    \hat{I}^i_k = \frac{I^i_k\,\text{tr}\bar{\bI} - \bar{X}_1^2\,\int\limits_\Omega c\,\bar{X}_2^2\,\text{d}V - \bar{X}_2^2\,\int\limits_\Omega c\,\bar{X}_1^2\,\text{d}V+ 2\bar{X}_1\bar{X}_2\,\int\limits_\Omega c\,\bar{X}_1\bar{X}_2\,\text{d}V}{2 I^i_k - \text{tr}\bI},
    \label{equ:Ihat}
\end{equation}
thus introducing a simple mechanical model that penalizes shape changes of the cell. When combined with the governing parabolic partial differential equation in conservation form \eqref{equ:dynamic}, and the boundary condition $\kappa \grad c_k \cdot \bn = 0$ on $\partial\Omega$ for $k = 1,\dots N$, we have a description for multi-cell growth, division and soft packing with the penalization of shape change. 

\subsection{Numerical implementation of the extended model}
The time discretized dynamics are now written as an explicit-implicit scheme. Given the initial conditions $\{c_k^0,\mu_k^0\}$ and the solution $\{c_k^n,\mu_k^n\}$, the time-discrete versions of Equations \eqref{equ:dynamic} and \eqref{equ:chempotmech} are,
\begin{align}
c^{n+1}_{k} &= c^{n}_{k} + \Delta t (M~\grad \cdot (\grad \mu^{n+1}_{k})+s_{k})\nonumber\\
\mu_k^{n+1}  &= f_{,c_k^{n+1}} -  \kappa \Delta c_k^{n+1} + \sum_{l\ne k}2\lambda  c_k^{n+1} {c_l^{n+1}}^2 - \sum_{i=1}^{\text{dim}} 2\delta_k^i \left({I_k^i} ^\text{ref} - I_k^i\right)\hat{I}^i_k.
\label{equ:anisotropy_time_discretization}
\end{align}
Note that the mechanical terms exerting control over the shape do not carry time instant superscripts in Equation \eqref{equ:anisotropy_time_discretization} for the sake of brevity. We have experimented with evaluating these terms at $t^{n+1}$ in a fully implicit method, as well as at $t^n$, in an explicit-implicit implementation. The complicated functional evaluations in Equations \eqref{equ:CayleyHamilton}-\eqref{equ:Ihat} make the fully implicit method notably more expensive than the explicit-implicit method, which has been used in the numerical examples of Section \ref{sec:anisotropy_study}.

\red{The weak form is stated as follows: Find $c^{n+1}_k \in \mathscr{S} = \{c \vert c \in \mathscr{H}^1(\Omega),  \grad c \cdot\bn = 0 \;\text{on}\; \partial\Omega \}$ and $\mu^{n+1}_k \in \mathscr{S} = \{\mu \vert \mu\in \mathscr{H}^1(\Omega)\}$ such that for all variations $w \in \mathscr{V} = \{w \vert w\in \mathscr{H}^1(\Omega)\}$ on $c_k$ and $v \in \mathscr{V} = \{v \vert v\in \mathscr{H}^1(\Omega)\}$ on $\mu_k$, respectively, the following residual equations  are satisfied:
\begin{align}
\int\limits_{\Omega}   w c_k^{n+1} ~dV
=&\int\limits_{\Omega}   (w  c_k^{n} - \grad w \cdot \Delta t M \grad \mu_k^{n+1} + ws_k) ~dV\nonumber\\
\int\limits_{\Omega}   v  \mu_{k}^{n+1}  ~dV 
=&\int\limits_{\Omega} ( v  f_{,c_k}^{n} + \grad v \cdot \kappa \grad c_k^{n} ) ~dV + \int\limits_\Omega v \sum_{k\ne l}2\lambda  c_k c_l^2  ~dV \nonumber\\
&- \sum_{i=1}^\text{dim} 2\delta_k^i \left({I_k^i} ^\text{ref}- I_k^i\right)\int\limits_\Omega v \hat{I}^i_k ~dV
\label{equ:anisotropy_weak} 
\end{align}
}
\subsection{Parametric study of the mechanical moduli}
In our numerical experiments with the above model for mechanical control of shape change in packing, we have found that the effective values of the mechanical moduli $\delta^i_k$ that impose control on cell shape fall within the range $\delta^i_k \in [1\times 10^4, 2\times 10^5]$. Thus, with $\delta^1_k \to 0$ and $\delta^2_k$ in the above range, the model produces elliptical single cells of increasing aspect ratio with the major axis corresponding to the smaller principal moment of inertia $I^2_k < I^1_k$. This parametric study appears in Figure \ref{fig:penalization}. Much larger values  $\delta^i_k > 2\times 10^5$ make the problem stiff, and convergence of the nonlinear solver becomes difficult. Other aspects of the numerical implentation remain the same as in Sections \ref{sec:chemo_numerical}-\ref{sec:source}. 


\subsection{Numerical studies of soft packing of cells with penalization of shape changes}
\label{sec:anisotropy_study}
\begin{figure}[ht]
	\centering
	\begin{subfigure}[h!]{0.35\textwidth}
		\includegraphics[width=\textwidth]{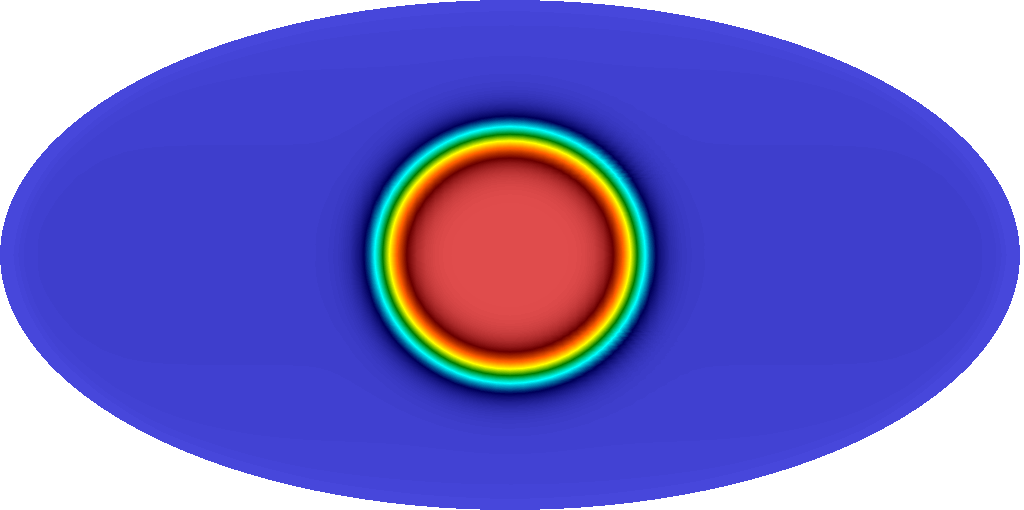}
		\caption{$\delta^2_1 = 50000$}
		\label{fig:delta50000}
	\end{subfigure}
	\begin{subfigure}[h!]{0.35\textwidth}
		\includegraphics[width=\textwidth]{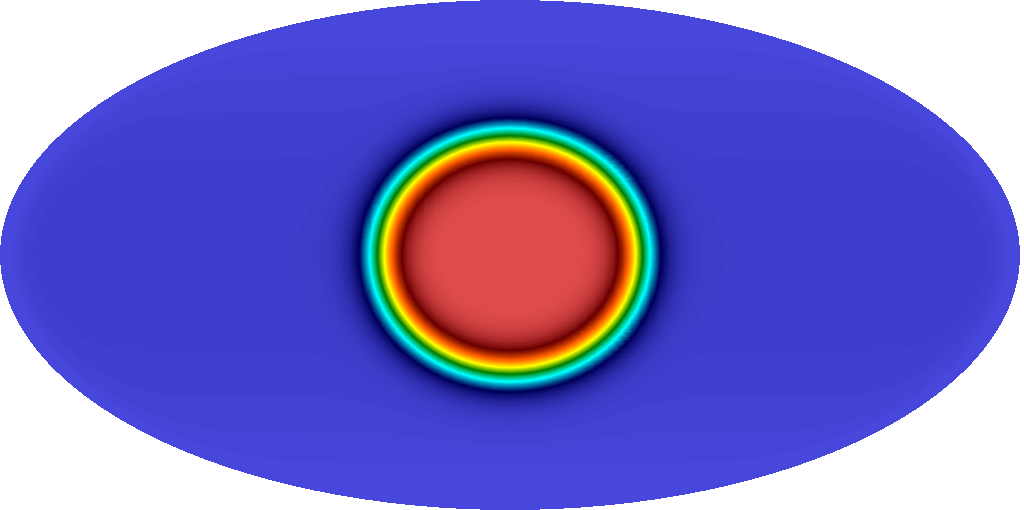}
		\caption{$\delta^2_1 = 100000$}
		\label{fig:delta100000}
	\end{subfigure}
	\begin{subfigure}[h!]{0.35\textwidth}
		\includegraphics[width=\textwidth]{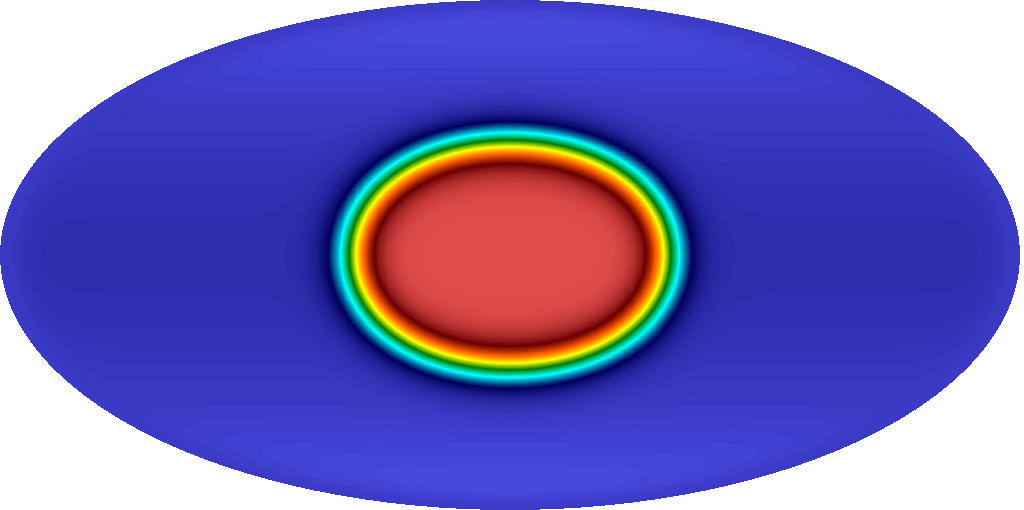}
		\caption{$\delta^2_1 = 200000$}
		\label{fig:delta200000}
	\end{subfigure}
	\caption{Fully developed single cell shape change due to anisotropic mechanical moduli $\delta^1_1 \to 0$ and $\delta^2_1 \in [1\times 10^4, 2\times 10^5]$.}
	\label{fig:penalization}
\end{figure}

In order to illustrate the control of cell shape  by the mechanical moduli $\delta^i_k$, we set $\delta^1_k \to 0$ and $\delta^2_k \in [1\times 10^4, 2\times 10^5]$. The results appear in Figure \ref{fig:penalization} for a single cell. The modulus $\delta^2_1$ takes on values $5\times 10^4$, $1\times 10^5$, and $2\times 10^5$ in Figures \ref{fig:delta50000}, \ref{fig:delta100000} and \ref{fig:delta200000}, respectively. It can be seen from these three figures that the larger mechanical modulus, $\delta^2_1$ constrains growth along the minor principal axis of the elliptical cell that forms. As $\delta^2_1$ becomes larger, it tends to produce an oblate shape, as the final equilibrium state in Figure \ref{fig:delta200000}. In comparison, at the lower value of $\delta^2_1 = 5\times 10^4$, this modulus does not much affect the ellipticity of cell shape, as seen in Figure \ref{fig:delta50000}. 

\begin{figure}[ht]
	\centering
	\begin{subfigure}[h!]{0.35\textwidth}
		\includegraphics[width=\textwidth]{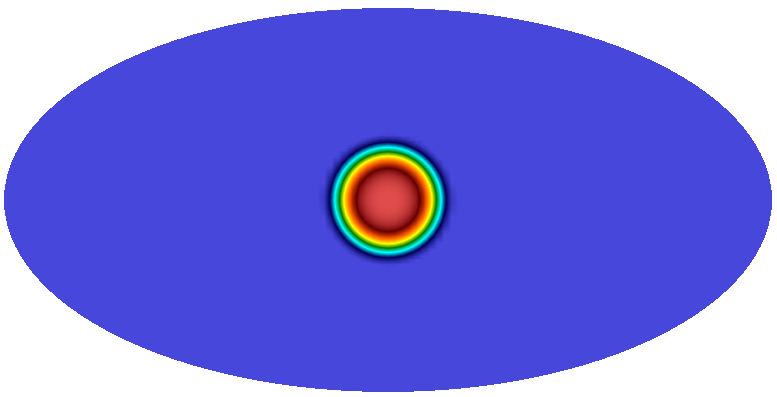}
		\caption{Mother cell}
		\label{fig:anisotropy_onecell}
	\end{subfigure}
	\begin{subfigure}[h!]{0.35\textwidth}
		\includegraphics[width=\textwidth]{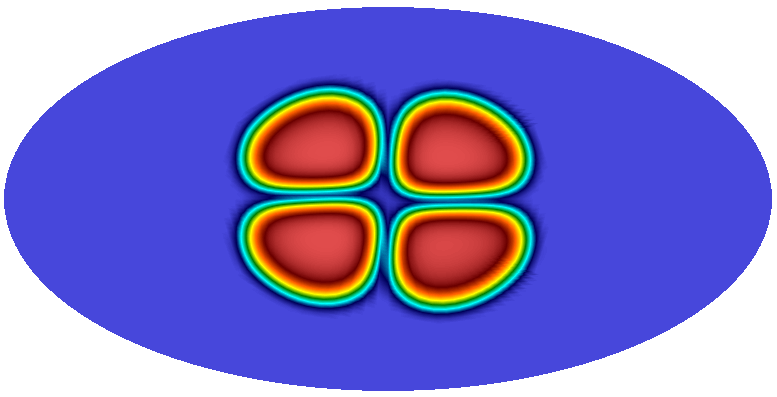}
		\caption{Four daughter cells}
		\label{fig:anisotropy_fourcell}
	\end{subfigure}
	\begin{subfigure}[h!]{0.35\textwidth}
		\includegraphics[width=\textwidth]{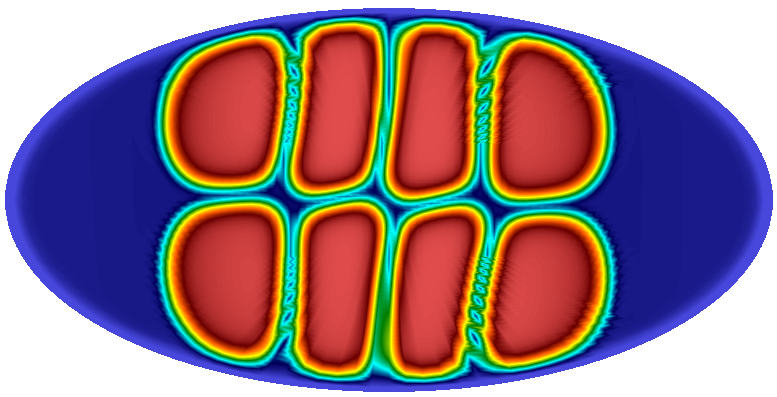}
		\caption{Eight daughter cells}
		\label{fig:anisotropy_eightcell}
	\end{subfigure}
	\begin{subfigure}[h!]{0.35\textwidth}
		\includegraphics[width=\textwidth]{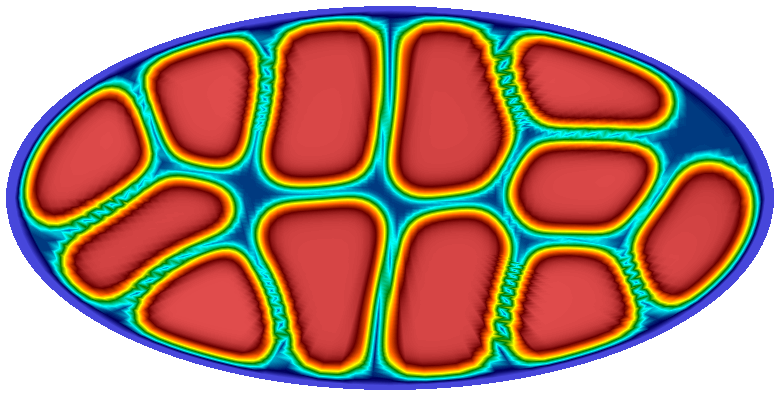}
		\caption{Twelve daughter cells}
		\label{fig:anisotropy_twelvecell}
	\end{subfigure}
	\caption{Cell division with penalization of cell shape change. The shapes of the 12-cell cluster differ from the 12-cell cluster in Figure \ref{fig:celldivision} due to the additional effect of mechanics. Cell interiors are shown in red and the cell membrane in cyan-yellow.}
	\label{fig:anisotropy_division}
\end{figure}

The progression of cell division from a single mother cell into twelve daughter cells is demonstrated in Figure \ref{fig:anisotropy_division}. Note the tight packing and shape changes attained by the cells. We also draw attention to the differences in cell shapes between Figures \ref{fig:celldivision} and \ref{fig:anisotropy_division}. This difference is especially notable at the 12-cell stage, and is  due to the added penalization of cell shape change already demonstrated in Figure \ref{fig:penalization}. We expect, also, that the shapes in Figure \ref{fig:anisotropy_division} are more physically accurate because they account for shape change, albeit by a simple mechanical model. 

\begin{figure}[ht]
\centering
\includegraphics[scale=0.5]{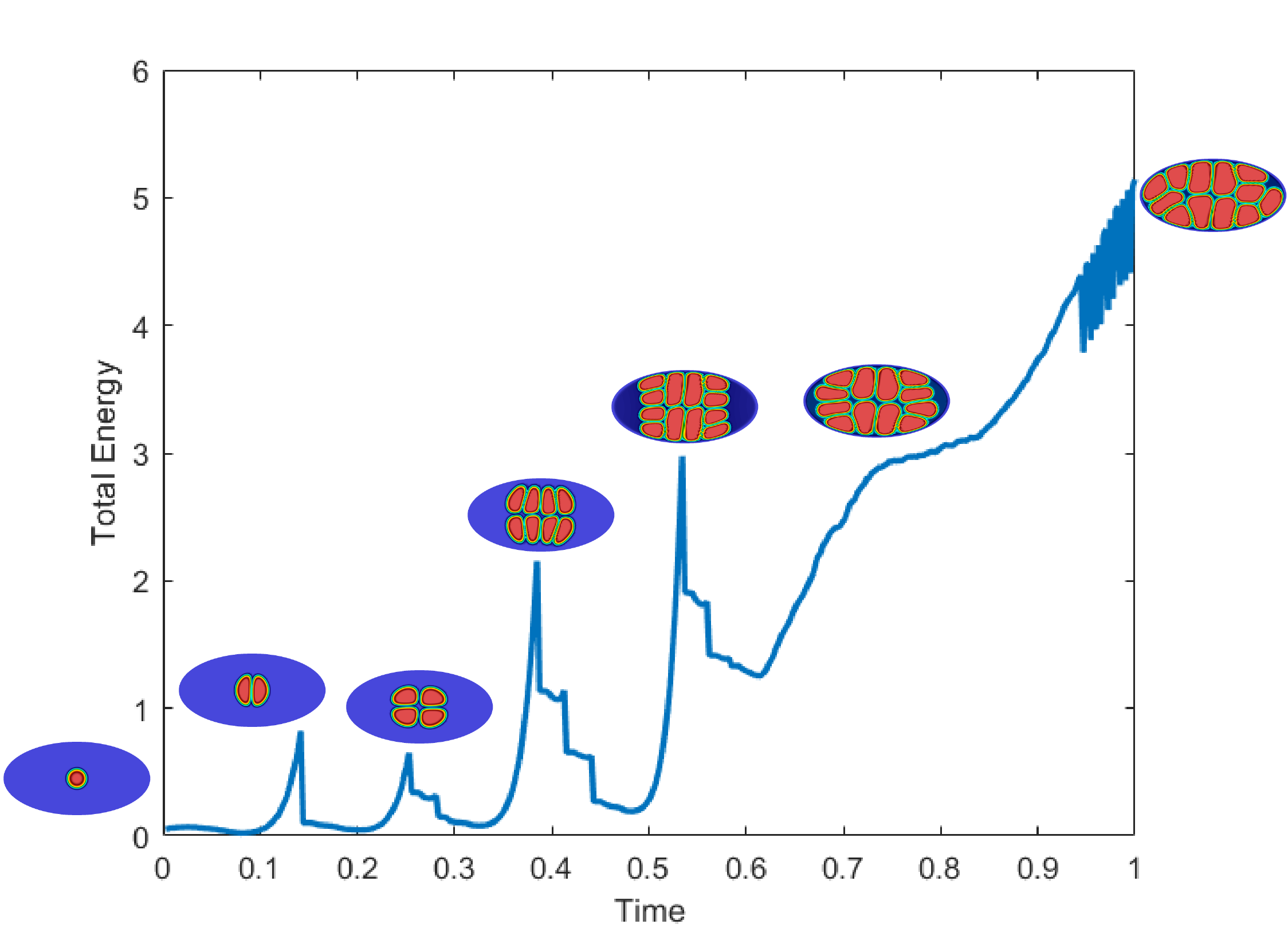}
\caption{Evolution of the total free energy with time (normalized). \red{Each spike in the energy curve corresponds to transient repulsion between newly formed daughter cells following a cell division, and the corresponding cell division events are shown in the inset sub-figures.}}
\label{fig:energy_anisotropy}
\end{figure}

The accompanying evolution of the total free energy is shown in Figure \ref{fig:energy_anisotropy}. As in Figure \ref{fig:energy}, transient fluctuations occur with each cell division due to the formation of a sharp boundary between the daughter cells, and the transiently stronger repulsion between the daughter cells. However, in this case, the transient fluctuations are more prominent and spread out due to the rapid changes in daughter cell shapes that themselves result from the elastic repulsion following division. Compared to Figure \ref{fig:energy}, the free energy values in Figure \ref{fig:energy_anisotropy} are much higher. This is due to the additional penalization in the form of the mechanics term whose relative magnitude has been made higher than the regular Cahn-Hilliard and overlap penalization terms. As observed before, the height of a spike and its width on the time axis increase with the number of cell divisions occurring in that time interval and the mass source causes the gradual increase in the free energy over time.

\begin{figure}[ht]
	\centering
	\begin{subfigure}[h!]{0.8\textwidth}
		\includegraphics[width=\textwidth]{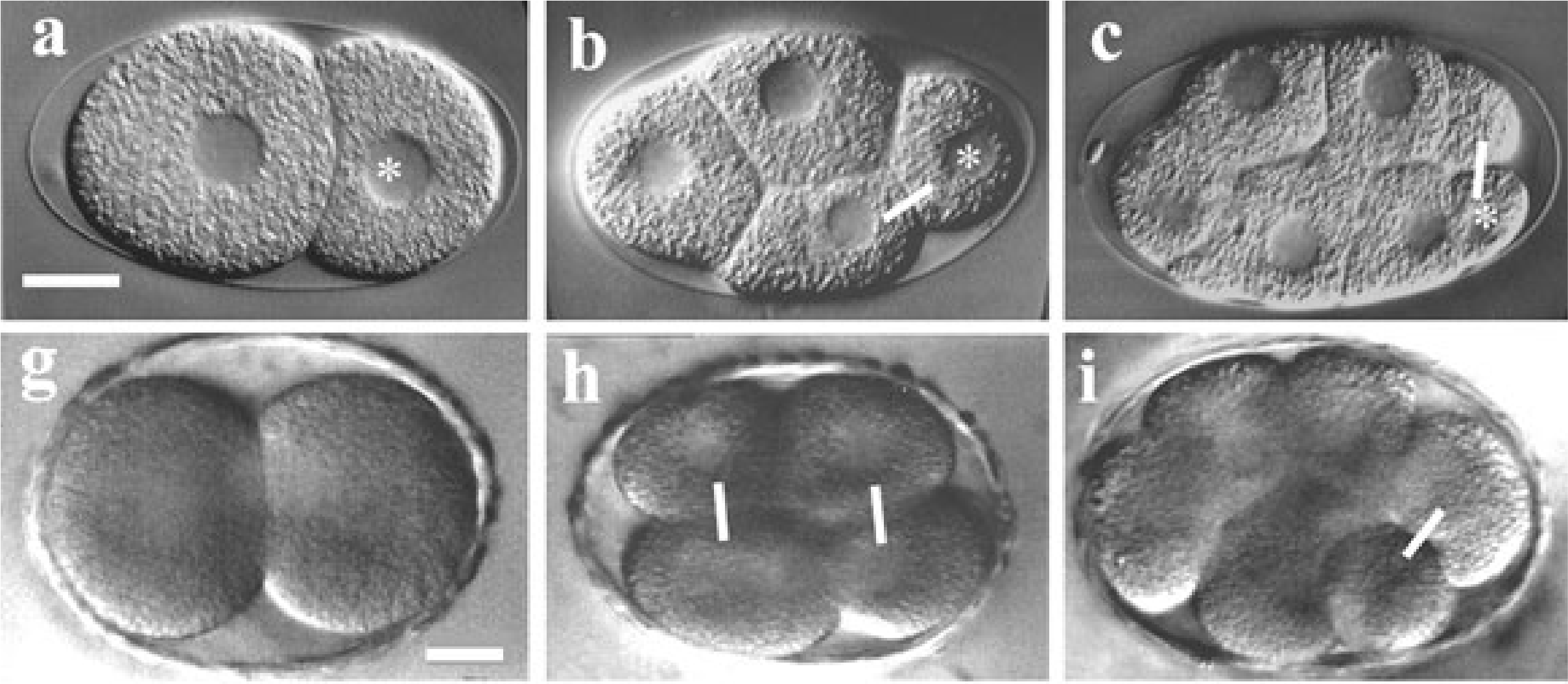}
		\caption{\red{Early embryogenesis of nematodes up to the eight-cell stage observed in \emph{C. elegans}, (a-c), and \emph{Prionchulus sp.}, (g-i). Some of the cells are either partially or completely hidden from view in (c) and (i). Figure reproduced from Schierenberg \cite{Schierenberg2006} (Original figure distributed under Creative Commons Attribution License).}}
		\label{fig:differentiable} 
	\end{subfigure}
	\begin{subfigure}[h!]{0.8\textwidth}
	    \vspace{0.1in}
		\includegraphics[width=\textwidth]{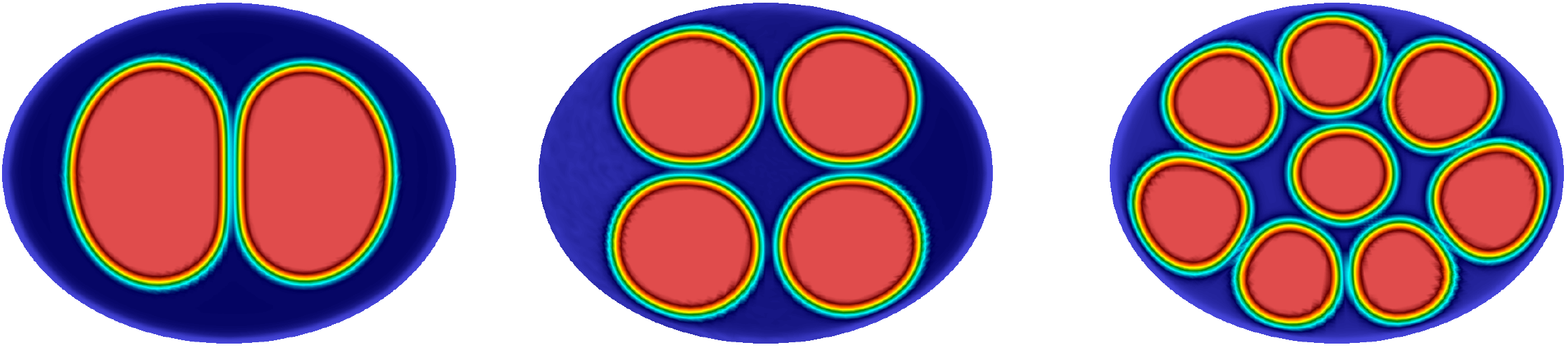}
		\caption{\red{Computations showing up to the third generation of divisions resulting in eight cells at a constant total cell volume.}}
		\label{fig:eightCell}
	\end{subfigure}
	\caption{\red{Comparison of cell morphologies during early embryogenesis observed in (a) nematodes (\emph{C. elegans} and \emph{Prionchulus sp.}), and (b) the computational model.}}
	\label{fig:embryogenesis}
\end{figure}

\section{Conclusion}
\label{sec:concl}
In this paper we have presented a phase field-based diffuse interface framework for modeling the \red{growth, division and packing} of multi-cell aggregates. The model allows for high fidelity representation of smooth, anisotropic and \red{unsymmetric} cell geometries, cell-cell contact and the resulting mechanical compaction processes intrinsic to soft packing in multi-cell aggregates. The cells are delineated by conserved scalar phase fields driven by the growth, division and compaction processes. The zero level set of the phase field representing a given cell also identifies the cell membrane. The driving force is the minimization of a free energy functional, and the governing equations are variationally derived to yield a system of parabolic partial differential equations. The salient features of the formulation are: 
\begin{itemize}
\item Being a field formulation, no discreet interface evolving mechanisms are needed, such as those employed in lattice, cell-centric and vertex dynamics models. 
\item \red{The dynamics occurs at time scales controlled by physically meaningful mechanisms: the growth rate and the doubling time of cells. Absent are severely reduced (by orders of magnitude) time steps needed to equilibrate the very high frequency response of vertices in, e.g., cellular automata models. We note also that such, very high frequency response, is not physically realistic in the highly dissipative setting of cell biophysics. On the other hand, the first-order dynamics of phase field models does impose dissipation that can be tuned to the relevant timescale.}
\item The cell boundary is represented by a continuously differentiable zero level set of the phase field and can thus represent general, smooth cell shapes without being limited to polygons or a jagged representation of the cell boundary. 
\item The mechanics of soft packing is modelled by penalizing departures of the principal moments of inertia of each cell from their initial values. This is a very simplified representation of the anisotropic mechanical response of the underlying cytoskeleton and cell membrane.
\end{itemize}
\red{These advantages significantly differentiate our model from cellular automata based cell-centric and vertex dynamics models used in the popular frameworks for modelling cell populations such as Cell-based Chaste~\cite{chaste2013} and CompuCell3D~\cite{compuCell2012}. However, the increased accuracy of cell shapes and mechanics does come at a higher computational cost needed to solve coupled partial differential equations over the problem domain. The targets of this numerical framework are applications requiring an accurate representation of cell shapes and modeling of long range mechanical interactions. In biological processes involving packing of cells in a confined space, interesting spatial differentiation/patterning of cell clusters are observed (e.g. early stages of Zygote-Morula-Balstocyst-Gastrula evolution in embryonic development \cite{itskovitz2000}, stem cell heterogeneity in tumors \cite{reya2001} and tumor shape evolution \cite{mills2014, rudraraju2013}). Intrinsic to the process of packing is the effect of mechanical interactions of the cells among themselves and with the extra cellular environment. These phenomena operate on length scales comparable to the dimensions of the cells, and also introduce the effect of cell shape in the packing dynamics. A case in point is the early stage of embryogenesis (Zygote to the Morula and the Blastocyst) that involves the growth of a cell cluster beginning with a single cell that divides and grows to a few hundred. An attempt at modelling some of these processes (cleavage and compaction) in the early stage of embryogenesis using this framework is shown in Figure~\ref{fig:embryogenesis}.  The cell packing morphology predicted by our computational framework is compared to experimentally observed embryo morphology patterns in nematodes. As can be seen, the close correlation of cell volumes, shapes and their spatial distribution is encouraging. \\ \\ 
Currently, the framework does not include a model for cell migration and only has a very simplified representation of cell-cell mechanical interactions through the penalization of cell shape. However, migration can be modelled by incorporating an advection term. The authors have demonstrated such models previously for stress-driven movement of tumor stem cells \cite{rudraraju2013}, and recently for neuronal migration in the developing brain \cite{verner2018}. We note, also, that the simplified mechanics model, when applied to the shape changes of a sphere into an arbitrary ellipsoid, has reproduced the essential features of the neo-Hookean strain energy function in our studies (to be reported in a future communication).  Incorporation of pointwise constitutive models (e.g. general hyperelastic models, viscoelasticity and poroelasticity), required for a physically complete treatment of mechanics, also needs the formulation of the governing equations of mechanics in a fully Eulerian setting \cite{kamrin2012}. Such extensions will be incorporated into future developments of this framework. Lastly, the computational cost of the current model can be significantly reduced by removing the linear dependence of the number of degrees of freedom on the number of cells using the active parameter tracking method, as described in Section ~\ref{subsec:cellDivision}. This, also, will be presented in a future communication.
}

\bibliographystyle{plain}
\bibliography{references}

\begin{thebibliography}{10}

\bibitem{Cahn1979}
Samuel~M. Allen and John~W. Cahn.
\newblock A microscopic theory for antiphase boundary motion and its
  application to antiphase domain coarsening.
\newblock {\em Acta Metallurgica}, 27(6):1085 -- 1095, 1979.

\bibitem{Silvanus2017}
Silvanus Alt, Poulami Ganguly, and Guillaume Salbreux.
\newblock Vertex models: from cell mechanics to tissue morphogenesis.
\newblock {\em Philosophical Transactions of the Royal Society of London B:
  Biological Sciences}, 372(1720), 2017.

\bibitem{BangerthHartmannKanschat2007}
W.~Bangerth, R.~Hartmann, and G.~Kanschat.
\newblock {deal.II} -- a general purpose object oriented finite element
  library.
\newblock {\em ACM Trans. Math. Softw.}, 33(4):24/1--24/27, 2007.

\bibitem{Brezzi1991}
F.~Brezzi and M.~Fortin.
\newblock {\em Mixed and Hybrid Finite Element Methods}.
\newblock Springer-Verlag, 1991.

\bibitem{Brodland2004}
G.~W. Brodland.
\newblock Computational modeling of cell sorting, tissue engulfment, and
  related phenomena: A review.
\newblock {\em Applied Mechanics Reviews}, 57:47--76, 2004.

\bibitem{Cahn1958}
John~W. Cahn and John~E. Hilliard.
\newblock Free energy of a nonuniform system. i. interfacial free energy.
\newblock {\em The Journal of Chemical Physics}, 28(2):258--267, 1958.

\bibitem{FLETCHER20142291}
Alexander Fletcher, Miriam Osterfield, Ruth~E. Baker, and Stanislav~Y.
  Shvartsman.
\newblock Vertex models of epithelial morphogenesis.
\newblock {\em Biophysical Journal}, 106(11):2291 -- 2304, 2014.

\bibitem{Gilbert2000}
S.F. Gilbert.
\newblock {\em Developmental Biology. $6^{th}$ edition.}
\newblock Sinauer Associates, 2000.

\bibitem{Glazier1993}
James~A. Glazier and Francois Graner.
\newblock Simulation of the differential adhesion driven rearrangement of
  biological cells.
\newblock {\em Phys. Rev. E}, 47:2128--2154, Mar 1993.

\bibitem{GOEL1970423}
Narendra Goel, Richard~D. Campbell, Richard Gordon, Robert Rosen, Hugo
  Martinez, and Martynas Yaas.
\newblock Self-sorting of isotropic cells.
\newblock {\em Journal of Theoretical Biology}, 28(3):423 -- 468, 1970.

\bibitem{GOEL1978103}
Narendra~S. Goel and Gary Rogers.
\newblock Computer simulation of engulfment and other movements of embryonic
  tissues.
\newblock {\em Journal of Theoretical Biology}, 71(1):103 -- 140, 1978.

\bibitem{GRANER1993455}
Francois Graner.
\newblock Can surface adhesion drive cell-rearrangement? part i: Biological
  cell-sorting.
\newblock {\em Journal of Theoretical Biology}, 164(4):455 -- 476, 1993.

\bibitem{Glazier1992}
Francois Graner and James~A. Glazier.
\newblock Simulation of biological cell sorting using a two-dimensional
  extended potts model.
\newblock {\em Phys. Rev. Lett.}, 69:2013--2016, Sep 1992.

\bibitem{Trilinos2003}
Michael Heroux, Roscoe Bartlett, Vicki Howle~Robert Hoekstra, Jonathan Hu,
  Tamara Kolda, Richard Lehoucq, Kevin Long, Roger Pawlowski, Eric Phipps,
  Andrew Salinger, Heidi Thornquist, Ray Tuminaro, James Willenbring, and Alan
  Williams.
\newblock {An Overview of Trilinos}.
\newblock Technical Report SAND2003-2927, Sandia National Laboratories, 2003.

\bibitem{HONDA1978523}
Hisao Honda.
\newblock Description of cellular patterns by dirichlet domains: The
  two-dimensional case.
\newblock {\em Journal of Theoretical Biology}, 72(3):523 -- 543, 1978.

\bibitem{HONDA1983191}
Hisao Honda.
\newblock Geometrical models for cells in tissues.
\newblock {\em International Review of Cytology}, 81:191 -- 248, 1983.

\bibitem{Honda1986}
Hisao Honda, Hachiro Yamanaka, and Goro Eguchi.
\newblock Transformation of a polygonal cellular pattern during sexual
  maturation of the avian oviduct epithelium: computer simulation.
\newblock {\em Development}, 98(1):1--19, 1986.

\bibitem{itskovitz2000}
Joseph Itskovitz-Eldor, Maya Schuldiner, Dorit Karsenti, Amir Eden, Ofra
  Yanuka, Michal Amit, Hermona Soreq, and Nissim Benvenisty.
\newblock Differentiation of human embryonic stem cells into embryoid bodies
  compromising the three embryonic germ layers.
\newblock {\em Molecular medicine}, 6(2):88, 2000.

\bibitem{kamrin2012}
K.~Kamrin, C.~H. Rycroft, and J.~C. Nave.
\newblock Reference map technique for finite-strain elasticity and fluid--solid
  interaction.
\newblock {\em Journal of the Mechanics and Physics of Solids},
  60(11):1952--1969, 2012.

\bibitem{li05}
Xiaoye~S. Li.
\newblock An overview of {SuperLU}: Algorithms, implementation, and user
  interface.
\newblock {\em ACM Transactions on Mathematical Software}, 31(3):302--325,
  September 2005.

\bibitem{mills2014}
K.~L. Mills, R.~Kemkemer, S.~Rudraraju, and K.~Garikipati.
\newblock Elastic free energy drives the shape of prevascular solid tumors.
\newblock {\em PloS one}, 9(7):e103245, 2014.

\bibitem{chaste2013}
Gary~R Mirams, Christopher~J Arthurs, Miguel~O Bernabeu, Rafel Bordas, Jonathan
  Cooper, Alberto Corrias, Yohan Davit, Sara-Jane Dunn, Alexander~G Fletcher,
  Daniel~G Harvey, et~al.
\newblock Chaste: an open source c++ library for computational physiology and
  biology.
\newblock {\em PLoS computational biology}, 9(3):e1002970, 2013.

\bibitem{Mochizuk1998}
Atsushi Mochizuki, Naoyuki Wada, Hiroyuki Ide, and Yoh Iwasa.
\newblock Cell-cell adhesion in limb formation, estimated from photographs of
  cell sorting experiments based on a spatial stochastic model.
\newblock {\em Developmental Dynamics}, 211(3):204--214, 1998.

\bibitem{Mosaffa2015}
Payman Mosaffa, Nina Asadipour, Daniel Mill{\'a}n, Antonio
  Rodr{\'i}guez-Ferran, and Jose J~Mu{\~{n}}oz.
\newblock Cell-centred model for the simulation of curved cellular monolayers.
\newblock {\em Computational Particle Mechanics}, 2(4):359--370, Dec 2015.

\bibitem{Munoz2017}
Payman Mosaffa, Antonio Rodríguez‐Ferran, and José~J. Muñoz.
\newblock Hybrid cell‐centred/vertex model for multicellular systems with
  equilibrium‐preserving remodelling.
\newblock {\em International Journal for Numerical Methods in Biomedical
  Engineering}, 34(3):e2928, 2017.

\bibitem{Nonomura2012}
Makiko Nonomura.
\newblock Study on multicellular systems using a phase field model.
\newblock {\em PLOS ONE}, 7(4):1--9, 04 2012.

\bibitem{reya2001}
T.~Reya, S.~J. Morrison, M.~F. Clarke, and I.~L. Weissman.
\newblock Stem cells, cancer, and cancer stem cells.
\newblock {\em nature}, 414(6859):105, 2001.

\bibitem{rudraraju2013}
S.~Rudraraju, K.~L. Mills, R.~Kemkemer, and K.~Garikipati.
\newblock Multiphysics modeling of reactions, mass transport and mechanics of
  tumor growth.
\newblock In {\em Computer Models in Biomechanics}, pages 293--303. Springer,
  2013.

\bibitem{Schierenberg2006}
E.~Schierenberg.
\newblock {\em Embryological variation during nematode development (January 02,
  2006), WormBook, ed. The C. elegans Research Community.}
\newblock WormBook, 2006.

\bibitem{compuCell2012}
Maciej~H Swat, Gilberto~L Thomas, Julio~M Belmonte, Abbas Shirinifard, Dimitrij
  Hmeljak, and James~A Glazier.
\newblock Multi-scale modeling of tissues using compucell3d.
\newblock In {\em Methods in cell biology}, volume 110, pages 325--366.
  Elsevier, 2012.

\bibitem{vedantam2006}
Srikanth Vedantam and BSV Patnaik.
\newblock Efficient numerical algorithm for multiphase field simulations.
\newblock {\em Physical Review E}, 73(1):016703, 2006.

\bibitem{verner2018}
S.N. Verner and K.~Garikipati.
\newblock A computational study of the mechanisms growth-driven folding
  patterns on shells, with application to the developing brain.
\newblock {\em Extreme Mechanics Letters}, 18:58--69, 2018.

\end{thebibliography}
\end{document}